# The impact of pressure on the critical miscibility of *n*-alkanes mixtures

Aleksandra Drozd-Rzoska, *Sylwester J. Rzoska, Jakub Kalabiński

Institute of High Pressure Physics Polish Academy of Sciences,

ul. Sokołowska 29/37, 01-142 Warsaw, Poland

(*) Corresponding author e-mail: sylwester.rzoska@unipress.waw.pl

**Abstract:**

Pressure evolutions of the critical consolute temperature $T_C$ in nitrobenzene, *o*-nitrotoluene and 1-nitropropane plus *n*-alkanes ($5 < n < 20$) mixtures are discussed. The crossover $dT_C/dP < 0 \to dT_C/dP > 0$ is revealed for mixtures containing nitrobenzene and *o*-nitrotoluene. For 1-nitropropane – *n*-alkanes mixtures the same $T_C(P)$ changes pattern, with $dT_C/dP > 0$, occurs. The explanation of such behavior is supported by the volume excess and nonlinear dielectric effect (NDE) studies. The parameterization for all $T_C(P)$ dependencies is proposed and successfully implemented.

The supplementary analysis of the coexistence curve under high pressure (nitrobenzene-decane, $P = 105 MPa$) confirmed suggestions of the isomorphism postulate of critical phenomena, but with the pretransitional anomaly of the diameter of the binodal notably stronger than observed under atmospheric pressure. The analysis within the homologous series of critical mixtures revealed dependencies for the evolution of the critical consolute concentration and temperature: $x_C(n) \propto \sqrt{n}$ and $T_C(n) \propto n^2$.

The critical opalescence is the classic phenomenon observed when approaching the critical point in liquids. This report shows the evidence for its unique 'initial phase': the critical opalescence in blue.

High-resolution experimental data under pressure were obtained via dielectric constant scans. They were possible due to the breakthrough design of the measurement capacitor enabling the simple and reliable isolation of tested samples from the pressurized medium.

**Introduction**

The *Physics of Critical Phenomena* is one of the grand successes of 20th-century science, due to the explanation of common patterns observed in microscopically distinct 'critical' systems.[1-5] It links the 'universal' pretransitional behavior on approaching the continuous phase transition to the dominance of multimolecular pretransitional fluctuations, which correlation length $\xi$ and lifetime $\tau$ boost on approaching the critical temperature $T_C$:[3,5]

$$\xi(T) = \xi_0 |T - T_C|^{-\nu} \qquad , \qquad \tau(T) = \tau_0 |T - T_C|^{-z\nu} \qquad (1)$$

where $\nu$ is the critical exponent of the correlation length and $z$ is the dynamical exponent. For critical mixtures of limited miscibility discussed below $\nu \approx 0.625$ and $z = 3$, what yields $z\nu \approx 1.875$; $\xi_0$ and $\tau_0$ are critical amplitudes.

On approaching $T_C$ different physical magnitudes are described by power-type relations with universal critical exponents, which values do not depend on microscopic features of systems but the space ($d$) and the order parameter ($n_{op}$) dimensionalities.[2-5] Cagniard de la Tour (1822) was the first who observed the consequence of Eqs. (1), as the unique 'milky' behavior in water, diethyl ether, and alcohols for some temperatures and pressures: the critical opalescence.[6,7] Thomas Andrews' research (1869) on near-critical liquid-gas phase equilibria in carbon dioxide (1869) linked these observations to the concept of the gas-liquid critical point.[8] Ten years later, Vladimir F. Alekseev described binary mixtures of limited miscibility for which the phase transition from the homogeneous liquid to two coexisting liquid liquid phases and the critical opalescence takes place.[9,10] The *Physics of Critical Phenomena* showed that the pretransitional behavior in the surrounding of the gas-liquid critical point (GLCP), the critical consolute point (CCP) of binary mixtures, simple magnetic systems with the paramagnetic - ferromagnetic transition and the critical point in 3D Ising model are within the same universality class ($d=3$, $n_{op}=1$). Critical exponents describing the precritical behavior are universal, i.e. they do not depend on microscopic features of a system but solely on $d$ and $n_{op}$.[2,3,5,10] For GLCP and CCP phase transitions are associated precritical fluctuations and the break of the translational symmetry when passing the critical temperature $T_C$. Describing this process order parameter (*op*) is linked to the shape of the coexistence curve. For binary mixtures of limited miscibility it is described via relations:[3,10]

$$\Delta x(T) = x_U - x_L = B(T_C - T)^\beta \left(1 + (T - T_C)^\Delta + ...\right) \tag{2}$$

$$d(T) = \frac{x_U + x_L}{2} = x_C + aT + A(T - T_C)^{1-\alpha}\left[1 + (T - T_C)^{\Delta_1} + ...\right] \tag{3}$$

where $\Delta x = x_L - x_U$ is the metric of the order parameter, $T < T_C$, $x_U$ and $x_L$ are related to the concentration of the selected component of the binary mixture in the upper and lower coexisting phase, respectively; $d(T)$ is for the diameter of the binodal. Values of the critical exponent for (3, 1) universality class: $\alpha \approx 0.115$ and $\beta \approx 0.325$. The term in the square bracket is associated with correction-to-scaling, important away from the critical point: $\Delta_1 \approx 0.5$ is the first correction-to-scaling exponent. For the gas-liquid transition, one can consider the density $\rho$ as the experimental base for the order parameter.

Linking Eqs. (2) and (3) one can describe the behavior of branches of the binodal (coexistence curve) $x_U(T)$ and $x_L(T)$. For decades the diameter (Eq. 3) was described via the Caillet-Mathew law of rectilinear diameter :[11] $d(T) = x_C + aT$ (CCP) or $d(T) = \rho_C + aT$ (GLCP), and served as the practical tool for determining the critical concentration or density.[12,13] Four decades ago it became evident that such simple description fails, but the experimental detection of *d(T)* precritical anomaly appeared to be the challenging experimental task due to its weakness matched with the experimental error.[3,10,12-14] Notable, that Eqs. (2) and (3) can linked to concentrations in mole fraction $x(T)$, volume fractions $\phi(T)$, ...[3,10] as well as such physical properties as density, refractive index, dielectric constant.[15,16] For the 'non-optimal' selection, additional power terms may appear.[3,10] One should stress, that the description of pretransitional effects via the single power term is possible only in the (very) immediate vicinity of the critical point.[1,3,5,10] In practice, the analysis of experimental data always requires some temperature range $\Delta T_{range}$, which leads to effective critical exponents. For instance for the order parameter:[10,17]

$$\Delta x(T) = x_U - x_L = B \Delta T^{\beta_{eff.}} \tag{4}$$

where the effective exponent: $\beta_{eff.} = \beta + \sum \Delta_i b_i \Delta T^{\Delta_i}$

Binary mixtures of limited miscibility with the critical consolute point constitute an important chapter of chemical physics, physics of phase transitions, and the soft matter physics.[1,3-5,18-21] They are also important for a variety of innovative applications ranging from chemical and material engineering,[22-27] mineral oils recovery and mineralogy[28-32] to biotechnology, foods and pharmacy technologies.[22,33] They benefit from unique properties in the surrounding of the critical consolute point, changes of properties when passing from the one- to the two-phase region and adjusting of mixtures components to the required specifications of the given process. However, in a one-component system there is only a single, isolated gas-liquid critical point.[1] For mixtures of limited miscibility, there is a line of critical point[3,16,22,34] and then the explanation of the impact of pressure on the system is essential. The postulate of isomorphism of critical phenomena[3] assumes parallel forms of pretransitional behavior, with the same critical exponents, for temperature and pressure path of approaching the critical point and for the temperature behavior under arbitrary pressure.[3] Surprisingly, the impact of pressure on the coexistence curve is still poorly evidenced. This covers the impact of pressure on the critical consolute temperature, where the most important is the phenomenological prediction via Myers-Smith- Katz- Scott (MSKS) relation:[35]

$$\frac{dT_C}{dP} = T_C \frac{V_E}{H_E} \tag{5}$$

where $H_E$ is the mixing enthalpy excess: for critical mixtures with the upper critical consolute point $H_E > 0$. $V_E$ is the volume excess of the mixture.

Notable, that the knowledge of the value of $dT_C/dP$ is essential for critical amplitudes of some properties tested under atmospheric pressure:[36-39]

$$E^+ \propto \left( \frac{2}{\varepsilon_0} \frac{dT_C}{dE^2} + \frac{dT_C}{dP} \right) \quad (6a)$$

$$C^+ = A^+ \frac{\rho_C}{\alpha(\alpha-1)} \frac{dT_C}{dP} \quad (6b)$$

$$R_\xi = \xi_0 \frac{\alpha A^+ \rho_C}{(dT_C/dP)k_B} \approx \nu \left( \frac{n_{op}}{4\pi} \right)^{1/d} \approx 0.271 \quad (6c)$$

where parameters are related to behavior for dielectric constant $\varepsilon(T) = \varepsilon_C + \varepsilon_b t + E^+ t^{1-\alpha} + ...$, specific heat $c(T) = c_C + c_b t + C^+ t^{-\alpha} + ...$, the thermal expansion coefficient $\alpha_P(T) = \alpha_C + \alpha_b + A^+ t^{1-\alpha} + ...$ and density $\rho(T) = \rho_C + \rho_b t + \Theta^+ t^{1-\alpha} + ...$, the correlation length $\xi(T) = \xi_0 t^{-\nu}$. They describe changes above the critical consolute temperature $T > T_C$ ($P = const$), as the function of $t = (T - T_C)/T_C$. $R_\xi$ is the universal 'critical' coefficient: note that it is determined solely by values of dimensionalities $d$ and $n_{OP}$, what is the key prediction for universal properties within the *Physics of Critical Phenomena.*

The question of the impact of pressure on the critical concentration also arises. To the best of the author's knowledge, this issue was addressed only in ref.[40], where the following empirical dependence was suggested:

$$\frac{x_C(P) - x_C}{x_C} = \frac{T_C(P) - T_C}{T_C} \quad (7)$$

where $x_C$ (mole fraction) and $T_C$ (K) are for reference data under atmospheric pressure. Eq.(7) was based on the analysis of experimental data for the low pressures regime, $P < 13 MPa$. Hence, the question of the influence of high pressures on the critical concentration remains.

Both the critical consolute temperature and concentration are non-universal properties, depending on the microscopic features of mixtures. Revealing some dependencies in this area is essential for fundamental modeling and practical implementations. One of possible ways of such insight can offer studies in homologous series of mixtures of limited miscibility. To the best of the author knowledge, extensive results are available only for polymer – low molecular weight solvents critical mixtures,[41-43] for which the universal behavior based on the mean-field theory by Flory et al.[44,45] is indicated:

$$T_C \approx \theta - 2\theta \times (N)^{1/2} \propto (N)^{1/2} \quad \text{and} \quad \phi_C \approx (N)^{-1/2} \tag{8}$$

where $N$ stands for the number of mono- units.

Yelash et al.[46] resumed the mentioned model and experimental values for $\phi_C(N)$ dependences and suggested the general function $\phi_C = \sqrt{P_2 + P_3 \ln(N)}/(P_1 + \sqrt{N})$ to explain the detected 'effective' description obtained in experiments: $\phi_C \propto N^r$, with $0.37 < N < 0.5$. Results recalled above are for the upper critical consolute point (UCCP). In refs.[47,48] the opposite case was tested, namely: (i) high molecular weight polystyrene - $n$-alkanes mixtures ( $5 < n < 13$), (ii) for such mixture the lower critical consolute point (LCCP) exists. The authors proposed the numerical solution for the Born-Green-Yvon (BGY) theory-based to test $T_C(n)$ behavior: the fair agreement for $8 < n < 13$ was obtained. Imre et al.[41] tested the impact of pressure on polystyrene ($M_W = 1.24$ kg/mol) plus $n$-alkanes (from octane to tetradecane) mixtures with the upper critical consolute point. They reported the shift $dT_C/dP < 0 \rightarrow dT_C/dP > 0$ for increasing length of $n$-alkanes. The parallel rise of $T_C(n)$ and critical concentration ($wt_C(n)$, weight fraction) were reported.

This report focuses on cognitive gaps for low-molecular-weight liquids critical mixtures indicated above. Studies were carried out in mixtures of nitrobenzene, $o$-nitrotoluene, and 1-nitropropane with $n$-alkanes, up to the high pressures regime. Experimental $T_C(P)$ dependences have been determined, and the effective way of their portrayal is proposed. The analysis of the coexistence curve under $P \approx 105 MPa$ confirmed the postulate of isomorphism of critical phenomena, but the 'amplification' of precritical effects is shown. Preliminary tests also revealed the new feature of the '*classic*' critical opalescence phenomenon: the initial state of the 'blue opalescence'. High-pressure studies were carried out via dielectric constant measurements, for which the novel design of the measurement capacitor was proposed.

## II. Experimental

Critical concentrations were estimated using the visual method, which is based on the analysis of the coexistence curve determined via observations of a set of ampoules with mixtures of different concentrations. They were placed in a transparent and thermostated set-up (50 L volume).[10,49,50] Observations started in the homogeneous liquid phase and on cooling the phase separation, associated with the binodal, appeared. Figure 1 shows the behavior for the mixture of the critical concentration ($x = x_C$, mole fraction) on cooling towards $T_C$. When passing $T_C$ the meniscus, separating coexisting phases, appears.

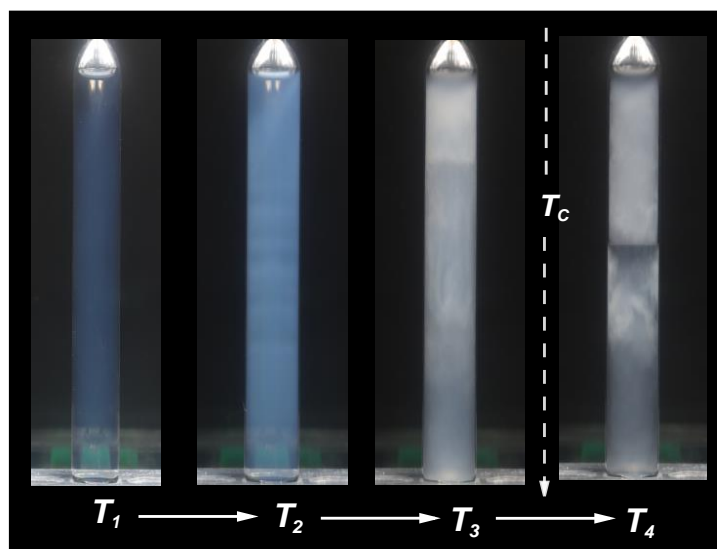

**Figure 1** The emergence of the critical opalescence on cooling from the homogeneous phase to the two-phase region in critical ($x = x_C$) nitrobenzene – tetradecane mixture of limited miscibility. The critical consolute temperature $T_C = 302.70K$. Photos are for $T_1 = T_C + 2.25K$, $T_2 = T_C + 0.95K$, $T_3 = T_C + 0.3K$, and $T_4 = T_C - 0.3K$.

The position of the meniscus does not change on further cooling. For non-critical mixtures the meniscus appears near the top or the bottom of the ampoule and notably shift when decreaing temperature. [50] In the critical concentration mixture cooling towards $T_C$ is signaled by the critical opalescence associated with notably different refractive indexes for critical fluctuations and the surrounded homogeneous liquid.[1,10] Close to $T_C$ tested samples were cooled down by $\Delta T = 0.1$ K steps and subsequently stabilized for at least 30 minutes (temperature stabilization $\Delta T = \pm 0.02$ K. For $T \gg T_C$ the mixture was transparent, but on cooling towards $T_C$ it first became slightly cloudy in deep-blue color ($T_1$ in Fig. 1). Subsequently, the picture changed to the more cloudy and light-blue pattern ($T_2$ in Fig.1). Finally, the mixture was profoundly cloudy and white/light grey ($T_3$ in Fig.1). When passing $T_C$ the meniscus appears near the half the height of the ampoule, but the critical opalescence reamined also in the two-phase region of coexisting liquid phases ($T_4$ in Fig. 4). To explain the color-related changes of the critical opalescence one can recall the classical Rayleigh relation for the intensity of the scattered light ($I_R$) by the inhomogeneous medium, often used for explaining the 'blue sky' origin: $I_R \sim 1/\lambda^4$. [51,52] Well above $T_C$ precritical fluctuations are small and short-living (Eq. (1)). Hence, the mostly scattered light is associated with the shortest wave-lengths (deep blue) component of the with light: this is "$T_1$" case. On approaching $T_C$ fluctuations grow up and into the scattered domain longer wavelength of the white light components can be included ($T_2$). In the immediate vicinity of the critical point there are all dimensions of critical fluctuations and all light wavelength are scattered ('white' $T_3$ and $T_4$ cases).

In the last decades, favorite features of soft matter have met the qualitative progress in high-pressure laboratory techniques.[22,53,54] The latter includes new, reliable and reusable systems of seals for high-pressure chambers, user-friendly and programmable systems for setting and controlling pressure and temperature. For high pressure *in situ* tests, particularly effective appeared to be the broadband dielectric spectroscopy (BDS), yielding insight into a variety of properties, from dynamics do intermolecular interactions. Such studies significantly advanced knowledge regarding the previtreous behavior of glass-forming system, which are mostly ultraviscous, almost semi-solid systems.[53] This causes that they are realtively resistive to possible contaminations from the pressurized liquid/medium, what makes the construction of the measurement capacitor placed within the high-pressure chamber relatively easy. For low molecular liquids based critical mixtures the sensitivity to contamination is enormous,[3] what leads to the requirement of the complete and reliable isolation between the tested sample and the pressurized liquid (Plexol in this research). In fact, probems associated with such end-module constitute the *'Achilles heel'* of high pressure studies in liquids.[22,53,54]

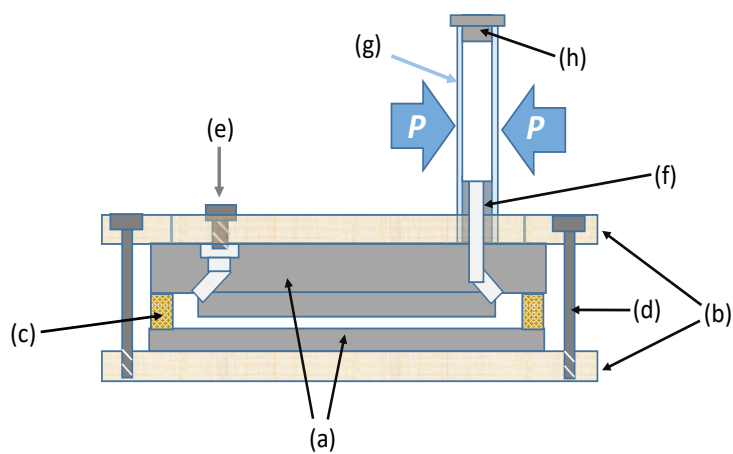

**Figure 2**  Capacitor for dielectric studies under pressure, offering the isolation between the sample and the pressurized medium: (**a**) capacitor plates (Invar, diameter $2r = 16$ mm), (**b**) clamp (PTFE, Teflon, PVDF, ..) (**c**) spacer between plates (the ring ca. 3 mm wide and 4-5 mm high; made from quartz, PTFE, ...), (**d**) screws linking clamps assembling plates of the capacitor, (**e**) the screw closing the 'channel' for introducing of the sample, (**f**) the sleeve for fixing the flexible/elastic tube (**g**) the elastic/flexible tube (Teflon or other elastic and resistant to the tested sample 'plastic') – it transmits pressure to the sample, (**h**) the stopper closing the flexible tube. There are 20 μm film washers (Teflon) between capacitor plates (a) and the spacer (c). For the screw (**e**) 1mm thick washer is required. Arrows with the letter '*P*' shows how pressure is transmitted to the sample

Results presented in this report applied observations of changes of dielectric constant appearing for the phase transition from the homogeneous liquid to the two-phase domain. For the first part of the studies, the measurement capacitor presented in refs.[39,55] was used . For this design, samples are in contact only with the stainless steel (Invar), quartz, and Teflon. The pressure is transmitted to tested samples via the deformation 20 $\mu$m thick Teflon film (membrane). The applied experimental solution enabled enables complete separation of the sample from the pressure transfer fluid (Plexol). Moreover, pressure could be increased and decreased without any contamination. Notwithstanding, the assembling was relatively difficult and the usage of the emmbrame was limited and often distorted by micro-pores in in the film transmitting pressure. The designed also requires relatively large diameter of the space within the pressure chamber (~35 mm). The new design of the measurement capacitor which lead to 100 % success experimental studies (what is not typical fro high pressure studies), free from mentioned limitations, is presented in Figure 2. The pressure is transmitted to the sample via the deformation of the flexible tube made from Teflon or any other elastic-plastic material, allowed for given studies. The elastic tube can be long and curled to occupy as little space as possible in the high-pressure chamber. The tube is multi-use, easily available, cheap, does not require preliminary preparations and and it is easy to replace. No parasitic leakages were observed even after twenty high-pressure experiments. Less than 1 cm$^3$ of a sample was required. After entering the two-phase region, the sample was heated or compressed/decompressed to return to the homogeneous liquid phase. It was additionally mixed via the sequence of high-voltage pulses from the apparatus for nonlinear dielectric effect (NDE) studies.[55] After the 'homogenization', the detection of the next pressure and temperature associated with the phase transition was detected. The pressure chamber was a part of the new IHPP PAS 'Unipress' system, offering the computer-controlled programming of pressure ($\pm$ 0.03 MPa, up to 900 MPa) and temperature ($\pm$0.02 K) in the temperature range between -30°C and +150 °C. The temperature was measured by means of the constantan - copper thermocouple placed in the body of the measurement capacitor The phase transition from the homogeneous liquid to the two-phase region was detected via the scan of dielectric constant (for the frequency $f = 10$ kHz), which value suddenly increased when passing the phase transition. High purity compounds were purchased from Sigma-Aldrich, but nitrobenzene ($C_6H_5NO_2$), 1-nitropropane ($C_3H_7NO_2$), and *o*-nitrotoluene ($C_7H_7NO_2$) were additionally distilled under vacuum, immediately before samples' preparation. Such treatment yielded samples with ca. two decades lower electric conductivity of vomposed than for 'just purchased' ones. Hydrocarbons (*n*-alkanes, which can be presented as the sequence of units H-$[CH_2]_n$-H, were used without additional purifications. The dielectric constant was monitoring via the *Alpha Novocontrol* impedance analyzer. Nonlinear dielectric effect (*NDE*) describes changes of dielectric constant caused by the impact of the strong electric field: $\varepsilon(E) = \varepsilon + \Delta\varepsilon E^2$ and then *NDE* is defined as $\Delta\varepsilon/E^2 = (\varepsilon(E) - \varepsilon)/E^2)/E^2$ .[55,56] The dual-field principle *NDE* experimental set-up was used.[77,85] The measurement capacitor was placed within the resonant circuit, and the tested sample interacted with the radio-frequency ($f = 3$MHz in the given case) weak measuring field ($U = 1$V, $d = 0.3$ mm and then $E = 0.44$kVm$^{-1}$). The strong electric field was applied via DC pulses ($U < 1500$ V)), lasting from $\Delta t = 1$ ms for *NDE* measurements and $\Delta t = 100$ ms for the supplementary homogenization of tested mixtures.

## III. Results and Discussion

### A. Critical consolute point

Studies were carried out for three homologous series of critical mixtures composed of nitro-compounds and *n*-alkanes. Figures 3 and 4 reveal simple dependences portraying changes of critical consolute temperatures and concentrations for these series:

$$[x_C(n)]^2 = x_o + a_n \times n \quad \rightarrow \quad x_C(n) \propto n^{1/2} \tag{9}$$

$$T_C(n) = T_o + b_n \times n^2 \quad \rightarrow \quad T_C(n) \propto n^2 \tag{10}$$

where $x_C$ is given in mole fractions and $T_C$ (K); $n$ stands for the number of carbon atoms in the molecule, or it is the metric of the length of $n$-alkanes molecules.

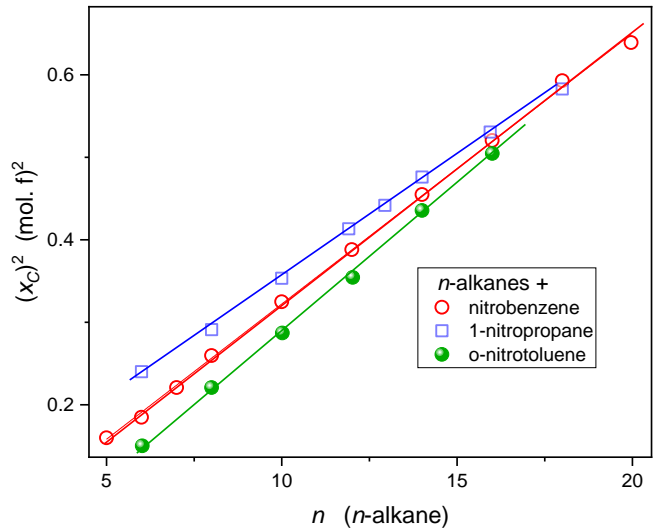

**Figure 3** The evolution of the critical concentration (mole fraction) for tested nitro-compounds + $n$-alkanes critical mixtures. The applied scale shows that for all tested homologous series $x_C(n) \propto \sqrt{n}$. Values of parameters related to Eq. (9) are given in Tab. I.

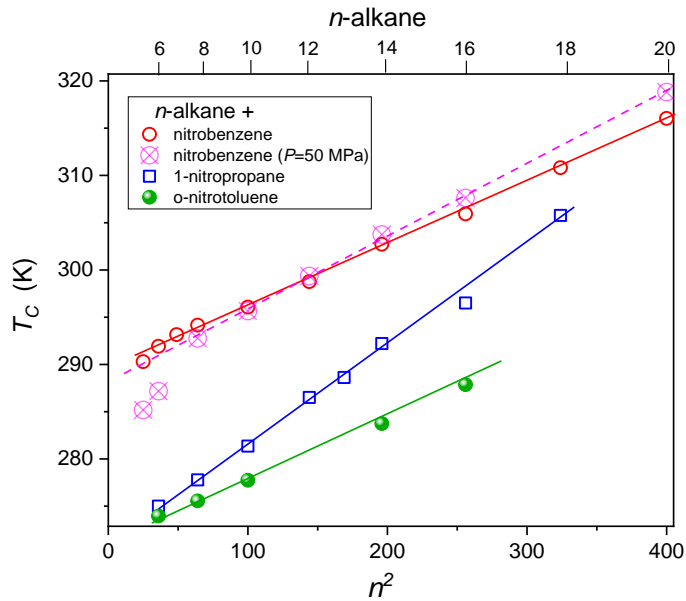

**Figure 4**  The evolution of the critical consolute temperature in nitro-compound – $n$-alkanes critical mixtures in the scale showing the prevalence for the description via the relation $T_C(n) \propto n^2$. Values of parameters related to Eq. (10) are collected in Table I.

**Table I**  Values of parameters for Eqs. (9) and (10), describing changes of the critical consolute temperature and concentration for tested homologous series of critical mixtures.

| Critical mixture | $x_0$ | $T_0$ (K) |
|---|---|---|
|  | $a_n$ | $b_n$ |
| nitrobenzene – | -0.0058 | 289.4 |
| $n$-alkanes, $P = 0.1$ MPa | 0.0328 | 0.0663 |
| nitrobenzene – | ------- | 285.6 |
| $n$-alkanes, $P = 50$ MPa |  | 0.086 |
| o-nitrotoluene – | 0.0656 | 271.5 |
| $n$-alkanes, $P = 0.1$ MPa | 0.036 | 0.065 |
| 1-nitropropane – | 0.0632 | 271.1 |
| $n$-alkanes, $P = 0.1$ MPa | 0.0292 | 0.105 |

Figures 5, 6 and 7 present evolutions of $T_C(P)$ in tested series of critical mixtures. The applied scale shows that for nitrobenzene and o-nitrotoluene + $n$-alkanes mixtures the transformation $dT_C/dP < 0 \rightarrow dT_C/dP > 0$, for rising $n$, takes place. For 1-nitropropane + $n$-alkanes mixture $T_C(P)$ experimental data overlap for all tested critical mixtures, and $dT_C/dP > 0$. As shown in Fig. 5 one can even 'construct' the quasi-critical mixture for which $dT_C/dP \approx 0$ in the broad range of pressure. This is associated with the '*fractional n-alkane*' defined as the mixtures of decane ($x = 0.6$-mole fraction) and undecane ($x = 0.4$-mole fraction) is used. One can consider it as the '*10.6 – alkane*'. All obtained $T_C(P)$ dependences can be well portrayed by the relation originally introduced for portraying pressure changes of the glass temperature, derived from the extended Clausius-Clapeyron equation:[57,58]

$$T_C(P) = k \times R(P) \times D(P) = k \times T_C^{ref} \cdot \left(1 + \frac{\Delta P}{\Pi}\right)^{1/b} \times exp\left(\frac{\Delta P}{c}\right) \qquad (11)$$

where $R(P)$ and $D(P)$ are for the rising and damping terms, respectively; $\Delta P = P - P_C^{ref.}$, $\Delta P = P - P_C^{ref.}$, $\Pi = \pi + P_C^{ref.}$, $-\pi$ is the extrapolated, negative pressure for which $T_C(P \to -\pi) \to 0$. The coefficient $k = +1$ if compressing increases $T_C$, for the opposite case $k = -1$. For the atmospheric pressure reference $\Delta P \approx P$.

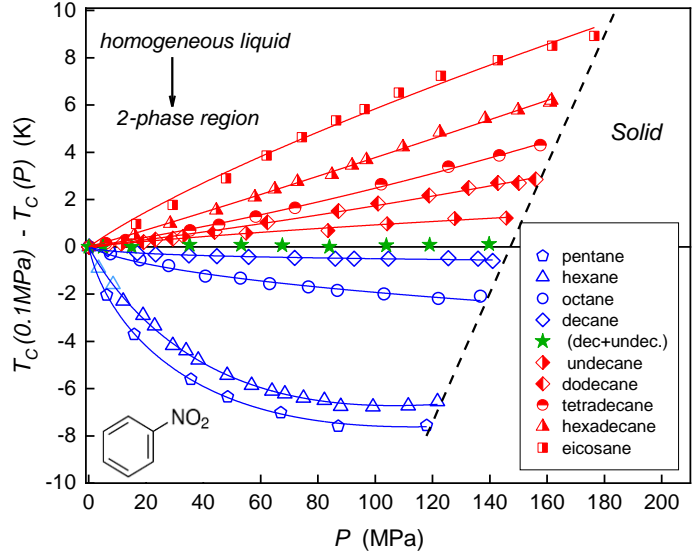

**Figure 5** The normalized pressure dependence of the critical temperature in nitrobenzene – *n*-alkanes critical mixtures. As the reference values of critical consolute temperatures under atmospheric pressure are taken. The structure of nitrobenzene molecule is shown. Solid curves show superior fitting of experimental data via Eq. (11), supported by the analysis via Eq. (12). Values of fitted parameters are collected in Table II.

Optimal values of parameters can be determined using the preliminary transformation of experimental data via dependence directly resulted from Eq. (11): [57,58]

$$\left(\frac{d\,lnT_C(P)}{dP}\right)^{-1} = b\Pi + b \times \Delta P = A + BP \tag{12a}$$

$$\left(\frac{d\,lnT_C(P)}{dP} + c^{-1}\right)^{-1} = b\Pi + b \times \Delta P = A + BP \tag{12b}$$

Plots $(d\,lnT_C/dP)^{-1}$ vs. $P$ yield the linear dependence if in Eq. (11) the description is related only to $R(P)$ term and $D(P) = 1$. Distortions indicate a possible extremum of $T_C(P)$ curve and the necessity of taking into account both $R(P)$ and $D(P)$ terms.. If the atmospheric pressure is taken as the reference one, can assume $P_C^{ref.} = 0$ in Eqs. (11) and then $\Delta P = P$, what yields $b = B$, $b = A/B$. Otherwise, values of $A$ and $B$ parameters are influenced by $P_C^{ref.} \neq 0$.

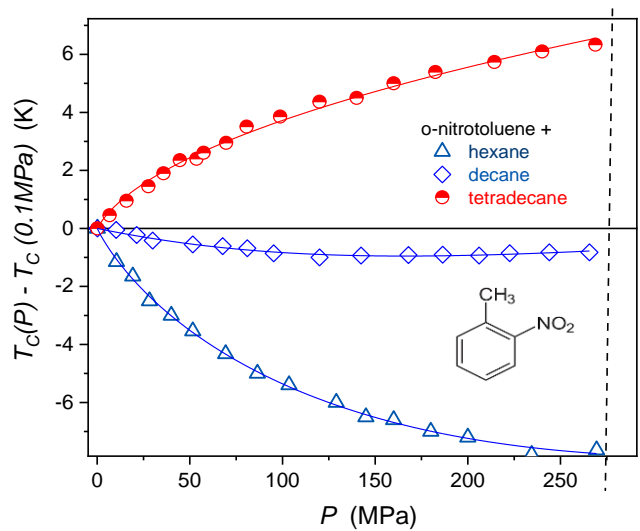

**Figure 6** Pressure dependences of normalized critical temperatures for o-nitrotoluene – n-alkanes critical mixtures. The structure of *o*-nitrotoluene is given. Solid lines are related to Eq. (11), with parameters given in Table II.

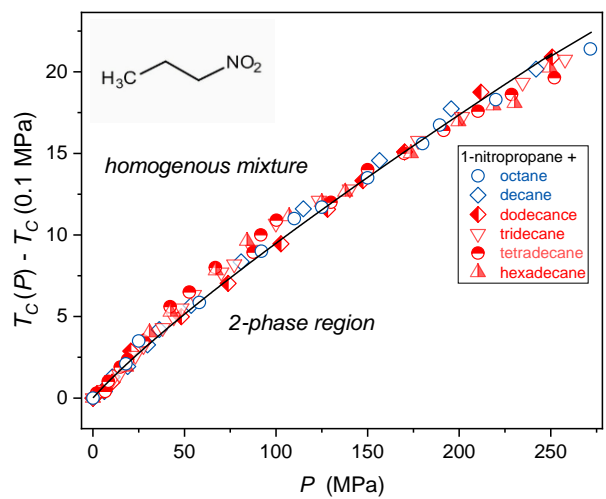

**Figure 7** The pressure evolution of critical temperatures for 1-nitropropane – n-alkanes critical mixtures. The structure of 1-nitropropane is shown. Solid lines are related to Eq. (11), with parameters given in Table II.

Figure 8 shows the example of the application of Eq. (12) for a selected critical mixture. The straight line shows the domain where $T_C(P) \propto R(P)$ in Eq. (11).

**Table II** Values of parameters for Eq. (11), portraying $T_C(P)$ experimental data in Figs. 5-7. As the reference, the critical consolute temperature under atmospheric pressure is taken.

| mixture | n-alkane | $T_C^{ref.}$ (K) | Π (MPa) | b | c (MPa) |
|---|---|---|---|---|---|
| nitrobenzene + n-alkane | eicosane | 316.05 | 8.8 | 1.3 | 0 |
| | hexadecane | 310.8 | 31.1 | 0.94 | 0 |
| | tetradecane | 302.7 | 83.8 | 0.72 | 0 |
| | dodecane | 298.9 | 103 | 0.65 | 0 |
| | undecane | 297.2 | 50 | 2.0 | 0 |
| | decane | 296.1 | 2.2 | 9.0 | 0 |
| | octane | 293.2 | 6.6 | 2.6 | 0 |
| | hexane | 291.9 | 3.4 | 1.23 | -0.008 |
| | pentane | 290.3 | 1.05 | 1.75 | -0.005 |
| o-nitrotoluene + n-alkanes | tetradecane | 303.7 | 4.34 | 2.05 | 0.002 |
| | decane | 296.6 | 72 | 1.6 | -0.001 |
| | hexane | 287.2 | 5.35 | 1.45 | -0.002 |
| 1-ntropropane + n-alkanes | from hexane (n=6) to octadecane (n-18) | n    $T_C$<br>6 - 274.9<br>8 - 277.8<br>10 - 281.4<br>12 - 286.5<br>13 - 288.6<br>14 - 292.2<br>16 - 296.5<br>18 - 305.8 | 6.65 | 1.2 | 0 |

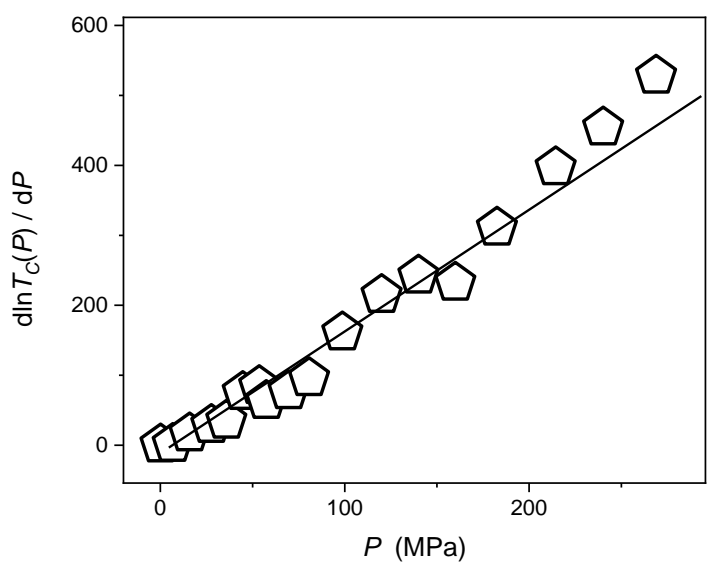

**Figure 8**  The derivative-based analysis (Eq. 12) showing the transformation of $T_C(P)$ experimental data focused on the description via Eq.(11).

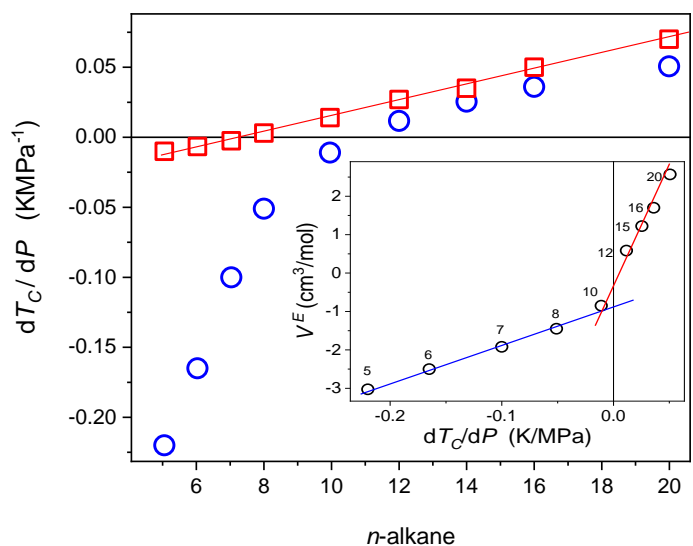

**Figure 9**  Changes of $dT_C/dP$ coefficient in nitrobenzene – *n*-alkanes critical mixtures. In the inset, they are presented as the function of the volume excess. Numbers are related to subsequent *n*-alkanes.

To discuss molecular origins of the obtained $T_C(P)$ behavior one can consider Figure 9, where the evolution of $dT_C/dP$ parameter for nitrobenzene – *n*-alkanes critical mixtures is presented. The inset shows fairly-well linear dependences $dT_C/dP$ vs. $V^E(n)$ (he volume excess). In fact, there are two linear dependences associated with critical mixtures characterized by $dT_C/dP < 0$ and $dT_C/dP > 0$. These results can be considered as the confirmation of the MSKS Eq. (5).

The discussion regarding the nature of presented above $T_C(P)$ behavior can also be supported by isothermal, concentrational studies of the nonlinear dielectric effect (NDE). In mixtures of unlimited miscibility, it is described by the Debye-Herweg-Piekara relation:[56,59]

$$\frac{\Delta \varepsilon}{E^2} = -F \frac{N_{dip.} \mu^4}{54 k_B T^3} R_S \qquad (13)$$

where $F(\varepsilon, \rho)$ is the local field factor, $N_{dip}$ denotes the number of permanent dipole moments in a unit volume, $\mu$ denotes the permanent dipole moment and $R_S$ stands for the dipole-dipole correlation factor.

Eq. (13) shows the extreme sensitivity of NDE to basic molecular properties, including intermolecular interactions.[56,59] The validation of Eq. (13) is shown in Fig. 10, where isothermal and concentration- related NDE behavior for slected mixtures is shown. For 1-nitropropane and CCl$_4$ mixture, the correlation factor $R_s(x) \approx 1$ (Eq. (13)) and the approximately linear decrease of NDE towards negative values occur on increasing the concentration of 1-nitropropane: $N_{dip.} \propto x$. The same pattern appears for diluted solutions of nitrobenzene and o-nitrotoluene with CCl$_4$. For higher concentrations of these nitro-compound compounds, the dipole-dipole coupling dominates what yields the increasingly negative correlation factor $R_s(x)$ and consequently the strongly positive NDE for $x \to 1$. The latter results from the coupling of benzene rings leading to the antiparallel dipole moments arrangements. Notable, that for nitrobenzene the coupling between benzene rings and the antiparallel arrangement of permanent dipole moments was confirmed experimentally and theoretically also beyond NDE studies.[60,61] Nitrobenzene – hexane is the only mixture for which limited miscibility presented in Fig. 10. The additional strong positive, contribution from pretransitional fluctuations appears in the vicinity of the critical concentration. For the almost-critical isotherm ($T \approx T_C$) one can describe changes of NDE as the critical effect:[55]

$$\left(\frac{\Delta\varepsilon}{E^2}\right)_C = \frac{\Delta\varepsilon}{E^2} - \left(\frac{\Delta\varepsilon}{E^2}\right)_{bckg} \propto |x-x_C|^{-\psi/\beta} \approx |x-x_C|^{-0.4/0.33} \approx |x-x_c|^{-1.2}, \quad T \approx T_C \qquad (14a)$$

The temperature behavior for the critical solution is given by [62,63]

$$\left(\frac{\Delta\varepsilon}{E^2}\right)_C \propto \chi_T \langle\Delta M^2\rangle_V \propto |T-T_C|^{-\psi} = |T-T_C|^{-(\gamma-2\beta)} \approx |x-x_c|^{-0.4}, \quad x \approx x_C \qquad (14b)$$

where $\langle\Delta M^2\rangle_V \propto (T-T_C)^{2\beta}$ denotes the average of local fluctuations of the order parameter and the susceptibility (compressibility) $\chi_T \propto (T-T_C)^{-\gamma} \approx (T-T_C)^{-1.23}$.

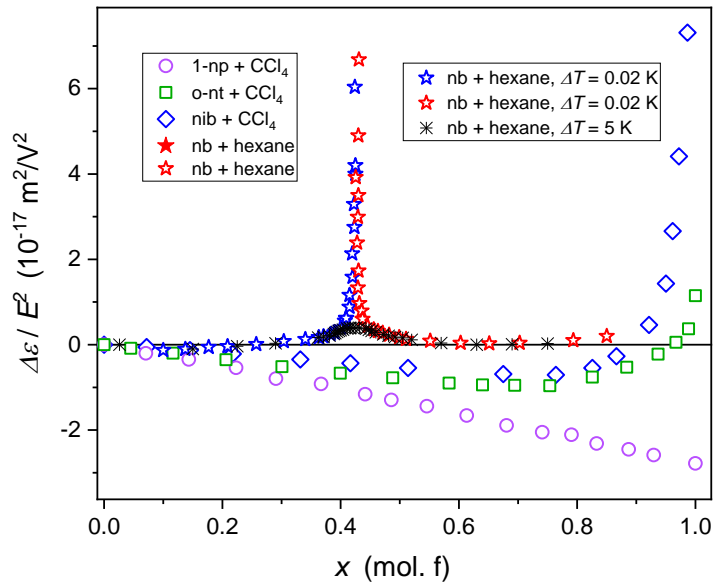

**Figure 10**  Nonlinear dielectric effect (*NDE*) in nitro-compound (nitrobenzene (nb), 1-nitropropane (1-np), *o*-nitrotoluene (*o*-nt)) solutions in a non-dipolar solvent (CCl$_4$, hexane). Studies in mixtures of unlimited miscibility, with CCl$_4$, were carried out for $T = 20^oC$. Nitrobenzene – hexane is the mixture of limited miscibility, with the critical concentration $x_C = 0.426$ mole fraction nitrobenzene: results for such mixtures are for the homogeneous liquid phase for $T = T_C + \Delta T$. Values of $\Delta T$ are given in the Figure.

Results above show that the impact of critical consolute point and critical fluctuations is 'propagated' deeply into the homogeneous liquid phase. Basing on the above discussion of the volume excess and *NDE* behavior one can assume that in mixtures containing 'short' *n*-alkanes (from pentane to decane), *n*-alkanes molecules may be 'hidden' between coupled nitrobenzene or *o*-nitrotoluene molecules. For longer *n*-alkanes, such mechanisms have to be absent. Notable, that all 1-nitropropane – *n*-alkane mixtures share the same pattern of $T_C(P)$ changes (Fig. 7). In the opinion of the authors, the question arises if this can be associated with the approximately uniaxial symmetry of both compounds and the slightly uniaxial ordering induced by isotropic compressing.

**B.  Coexistence curve**

The coexistence curve (binodal) determines the border between the homogeneous region of unlimited miscibility and the two- phase domain.[10] The primary way of describing coexistence curves under atmospheric pressure is the concentration vs. temperature plot.

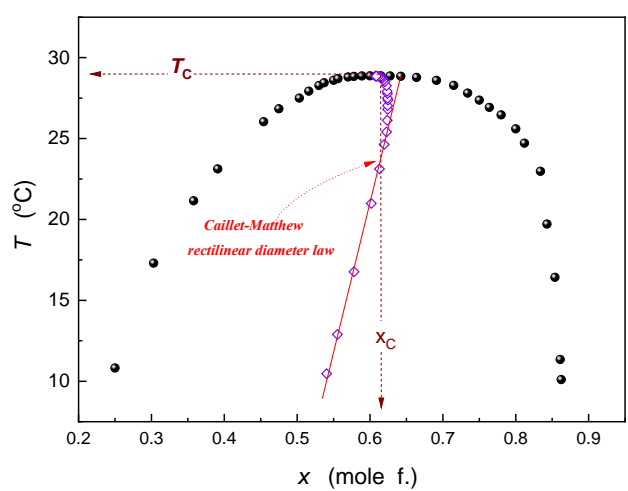

**Figure 11**   The coexistence curve for nitrobenzene – decane critical mixtures under $P = 105 MPa$. The straight line is for the law of rectilinear diameter.[57,58]

However, the impact of pressure on such binodals is still poorly evidenced.[3,4,10] This is undoubtedly due to the staff-requirements and difficulties associated with high pressure studies. Basically important is also the measuring module placed inside the high-pressure chamber, which must ensure the complete isolation from the pressure transmitting medium matched with high reliability. This is due to the unusual sensitivity and criticality in liquids for contaminations.[64-66] This fundamental problem has been solved by the new design of the measuring capacitor shown in Figure 1. Figure 11 presents the binodal for nitrobenzene – dodecane coexistence curve, based on dielectric constant measurements for several different concentration of mixtures ($x > x_C$ and $x < x_C$) determined under pressure $P = 105 MPa$.

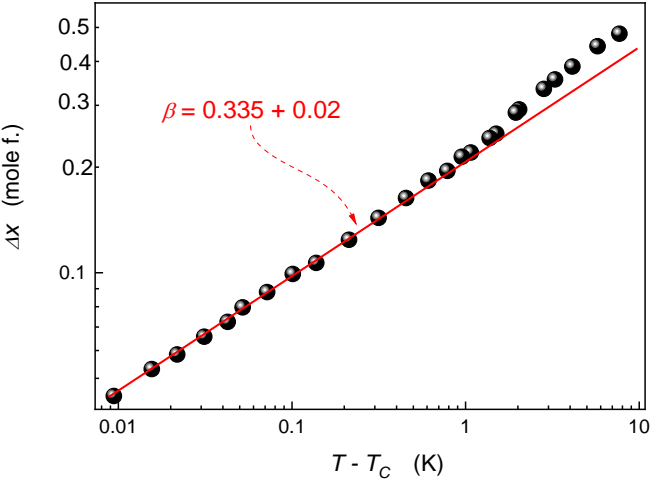

**Figure 12** The test for the critical, power type description (Eqs. (1 and 3)) of the order parameter for nitrobenzene – decane binodal under $P = 105 MPa$.

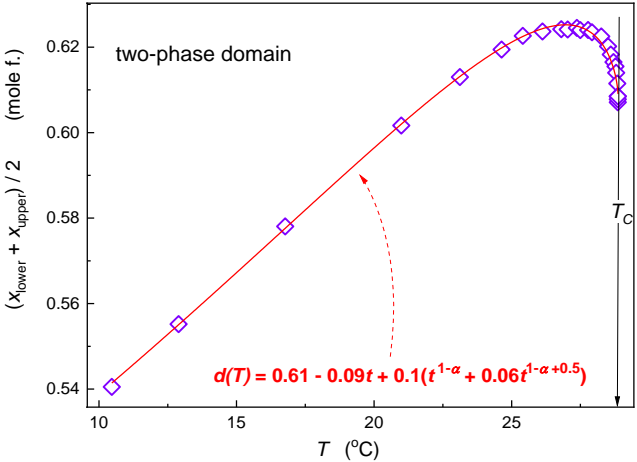

**Figure 13** The test for the critical, power type description via Eq. (2) for the diameter of the coexistence curve in for nitrobenzene – decane under pressure $P = 105 MPa$. Values of parameters are given in the Figure.

The analysis of the order parameter evolution based on these experimental data and Eqs. (2) and (4) is shown in Figure 12. It reveals the fairly well portrayal by the single power term (Eqs. (2) and (4)) with the critical exponent $\beta \approx 0.335$ for $T - T_C < 1K$, the value within the limit of the experimental error in agreement with the model one for $T \to T_C$,[1,5,3,10] $\beta \approx 0.325$. The diameter -focused plot presented in Figure 13 shows the fairly-well portrayal via Eq. (3), also remote from the critical point. Notable, that the manifestation of the pretransitional anomaly is significantly stronger than observed in studies under atmospheric pressure.[3,10] The inadequacy of the Caillet-Mathew law of rectilinear diameter is clearly shown in Fig. 11.[11,12] The analysis of the shape of the binodal showed a minimal shift of the critical concentration $\Delta x < 0.01$ mole fraction of nitrobenzene, in fair agreement with Jacobs Eq. (5).

## V. Summary

Results presented above showed the evidence regarding the impact of pressure on the critical consolute temperature in three homologous series of mixtures of limited miscibility composed of dipolar nitro-compounds and the non-dipolar solvent (*n*-alkanes). For nitrobenzene - *n*-alkanes and *o*-nitrotoluene – *n*-alkanes critical mixtures, the transformation $dT_C/dP < 0 \to dT_C/dP > 0$ is observed. The correlation of this phenomenon with the shift of the volume excess $V_E < 0 \to V_E > 0$ and the appearance of dimer intermolecular coupling for the dipolar component (NDE studies) is evidenced. Notable, that a similar shift $dT_C/dP$ can be noted for critical mixtures of oligostyrene – *n*-alkanes[41] and 1-alkanol – water[34], where the strong intermolecular coupling can also be expected. For 1-nitropropane and *n*-alkane critical mixtures, where the coupling is negligible, the same pattern of $T_C(P)$ evolution and $dT_C/dP > 0$ is shown. In agreement with general expectations of the physics of critical phenomena,[3] even high compressing has no impact on universal critical expponents characterizing the coexistence curve. Notwithstanding, the high compression causes the notable

'amplification' of the pretransitional anomaly of the diameterew, usually very weak under atmospheric pressure (Fig.13).[1,3,10] Finally worth stressing is the new evidence regarding changes of portrayals of $T_C(n)$ and $x_C(n)$ (Eqs. 9 and 10) within homologous series of tested compounds. .

The critical opalescence is amongst the 'founding stones' of the *Physics of Critical Phenomena*,[1,67] starting from observations of Cagniard de la Tour (1823)[6] and Thomas Andrews (1869)[8], and the grand theoretical contributions by Marian Smoluchowski (1908)[68] and Albert Einstein (1910)[69]. This report shows the evidence that there is an initaial '*blue opalescence state*' in critical mixtures of limited miscibility associated with the wavelengths – selective light scattering. Notable, that a message regarding a similar phenomenon, although in a very narrow range of temperatures, has been recently reported for the gas - liquid critical point in propane – hexane mixture.[70]

Concluding, this report presents a set of new dependences which can portray changes of the critical consolute temperature and concentration, as the function of pressure and within homologous series of low-molecular liquids composed of a nitro-compound and *n*-alkanes. It is shown that it is even possible to prepare a critical mixture for which $dT_C/dP \approx 0$ in a broad range of pressures. All these can be significant for fundamental modeling and for practical implementations exploring unique near-critical features.

**Acknowledgments**

Studies were carried out due to the support of the National Centre for Science (NCN OPUS grant, Poland), ref. UMO-2017/25/B/ST3/02458, the head: S. J. Rzoska.